# Transit Rider Heat Stress in Atlanta, GA under Current and Future Climate Scenarios

**Huiying Fan**
School of Civil and Environmental Engineering
Georgia Institute of Technology
790 Atlantic Dr, Atlanta, GA 30332
Email: fizzyfan@gatech.edu

**Geyu Lyu**
School of City & Regional Planning
Georgia Institute of Technology
790 Atlantic Dr, Atlanta, GA 30332
Email: glyu8@gatech.edu

**Hongyu Lu**
School of Civil and Environmental Engineering
Georgia Institute of Technology
790 Atlantic Dr, Atlanta, GA 30332
Email: hlu305@gatech.edu

**Angshuman Guin, Ph.D.**
School of Civil and Environmental Engineering
Georgia Institute of Technology
790 Atlantic Dr, Atlanta, GA 30332
Email: angshuman.guin@ce.gatech.edu

**Randall Guensler, Ph.D.**
School of Civil and Environmental Engineering
Georgia Institute of Technology
790 Atlantic Dr, Atlanta, GA 30332
Email: randall.guensler@ce.gatech.edu

**ABSTRACT**

Transit is a crucial mode of transportation, especially in urban areas and for urban and rural disadvantaged communities. Because extreme temperatures often pose threats to the elderly, members of the disability community, and other vulnerable populations, this study seeks to understand the level of influence that extreme temperatures may have on transit users across different demographic groups. In this case study for Atlanta, GA, heat stress is predicted for 2019 transit riders (using transit rider activity survey data) and for three future climate scenarios, SSP245, SSP370, and SSP585, into the year 2100. The HeatPath Analyzer and TransitSim 4.0 models were applied to predict cumulative heat exposure and trip-level risk for 35,999 trip equivalents for an average Atlanta area weekday in the summer of 2019. The analyses show that under 2019 weather conditions, 8.33% of summer trips were estimated to be conducted under extreme heat. With the projected future climate conditions, the percentage of trips under extreme heat risk grows steadily. By 2100, 37.1%, 56.1%, and 76.4% are projected to be under extreme heat risk for scenarios SSP245, SSP370, and SSP585, respectively. Under current weather conditions, Atlanta transit riders that own no vehicles and transit riders that are African American are disproportionately influenced by extreme heat. The disparity between these two groups and other groups of transit riders becomes wider as climate change continues to exacerbate. The findings of the study highlight an urgent need to implement heat mitigation and adaptation strategies in urban transit networks.

**INTRODUCTION**

With rapid urbanization and increasing frequency and severity of extreme temperature events, the thermal comfort of transit riders becomes more important for safe and reliable daily travel. Extreme heat is a common threat to urban citizens' health, particularly the elderly and people with special medical conditions [1]. In Atlanta, GA, extreme urban heat and humidity are among the most prevalent natural hazards, due to low latitude and a high percentage of impervious land [2]. Understanding traveller exposure to extreme temperatures throughout the transportation network is important for ensuring safe and comfortable travel.

While all travellers are affected by extreme temperatures, the level of impact across trips differs significantly. The use of transit modes, for example, typically exposes travellers to more outside conditions than the use of automobile modes. The vulnerability of specific groups of people is jointly determined by their exposure to extreme temperatures, their susceptibility to becoming adversely influenced, and their flexibility in choosing different travel alternatives. For example, historically, underprivileged neighbourhoods are particularly vulnerable to extreme temperatures due to higher reported exposure and lower access to urban greenery or other heat mitigation options at their household location [3–5]. In addition, underprivileged neighbourhoods tend to have much lower automobile access, reducing their flexibility in modal choice [6]. Senior citizens, with increased susceptibility to extreme temperatures [7,8], more frequently use transit and sidewalks, again due to reduced flexibility in mode choice [9]. Senior citizens also have reduced trip elasticity, or flexibility in making trip decisions [10,11], and are particularly vulnerable to extreme temperatures. Children, or travellers with children, are also more susceptible to extreme temperatures [12,13].

Despite increasing attention to climate change and extreme temperature events, uncertainty remains regarding how such events influence transit riders. This uncertainty not only arises from the complexity of climate systems but also the unpredictability of human behaviours such as trip decisions, mode choice, and route choice [14]. Recent research has started to explore the potential impacts of heat, rain, and wind on transit ridership [15], and some studies have incorporated exposure to extreme weather in assessing public transit accessibility [16]. However, few studies have extended their analysis into future scenarios to examine how transit riders might be affected and how these impacts may vary under different scenarios. This gap highlights the necessity for future prediction analyses to enhance understanding, helping transit agencies in making more informed decisions when planning for the future.

In this study, the HeatPath Analyzer and TransitSim 4.0 models work together calculate cumulative heat exposure and trip-level risk for each observed trip from a longitudinal transit onboard survey [17]. An equity assessment for different demographic groups, including income, race, age, gender, and vehicle ownership. Predictions for future years under three different climate scenarios are then intersected with the same demographic factors, to better understand the varying impacts of climate change across these groups into the future.

**DATA AND METHODS**

This section details the data and methods applied to the study. The first subsection introduces the study area and data used in the analyses, followed by a detailed description of the derivation of future meteorological conditions, cumulative exposure calculation methods, and the equity assessments. Appendix A provides a schematic overview of the methods employed in this study.

*Study area and data*

The Atlanta Metropolitan Area ("Metro Atlanta") is the major urban cluster in the State of Georgia. As of 2020, metro Atlanta was home to over six million people and encompassed nine counties. In addition to being the largest urban cluster, Metro Atlanta is also one of the most climate vulnerable regions in Georgia. Due to high percentage of impervious surfaces throughout the region, Metro Atlanta is prone to extreme heat [2]. Public transportation in metro Atlanta is operated by multiple transit agencies (service providers), with fixed-route transit options in five transit agencies (MARTA, Atlanta Streetcar, GRTA Express Bus, Gwinnett County Transit, and CobbLinc) (Figure 1).

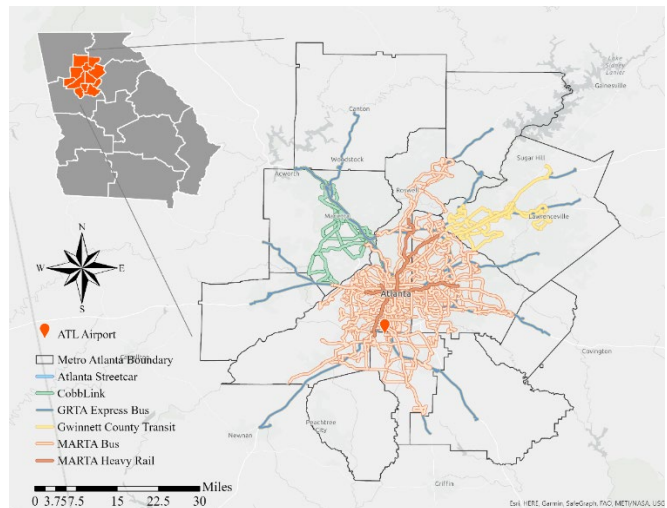

**Figure 1**. Study Area, Transit Agencies and Routes, and Location of ATL Airport (example network configuration during March 7th to April 18th, 2019)

The demographic data used in this study are derived from the Atlanta Regional Commission (ARC) Transit Onboard Survey ("the Survey"), which recorded 43,398 survey responses for travel across various routes from March to December 2019 [18]. The Survey is comprehensive and provides detailed demographic data for the transit riders. The survey data were all collected prior to the pandemic. Responses were weighted and expanded based on route, direction, time-of-day, and route segment, resulting in a dataset of 211,057 trip-equivalents that reflect the typical volume and composition of weekday trips in the region. After excluding 1,756 samples with incomplete data, 40,289 samples (95.8%) remained for the analysis, each including specific details such as origin, destination, and trip start time.

Data extracted from August 1st to September 15th, 2019 provide the activity for assessing extreme heat exposure during the summer. The selection was based on several factors. First, the Survey excluded June and July due to the school break, rendering the data non-representative of typical all summer conditions. The goal was to capture the period with the highest temperatures of the year. The chosen interval falls within the span from July 26th to September 29th, 2019, during which over 60% of the days saw maximum temperatures exceeding 90 degrees. In the end, the assessment employs 6,808 samples, representative of 35,999 trips on an average weekday.

*Projection of future meteorological condition*

High resolution meteorological information is needed for the study. Here, the analyses use hourly-binned meteorological information measured at the Hartsfield-Jackson Atlanta International Airport Weather Station (ATL, a central location to the study area, Figure 1). Methods used to extrapolate the data spatially are detailed in [19].

To assess future implications of exposure based on climate change, this research integrates future climate projections. Temperature changes were estimated using mean temperature change data from the U.S. Geological Survey's (USGS)'s National Climate Change Viewer (NCCV) dataset, which sources

CMIP6 estimates of the multi-model mean for Fulton County, GA [20]. This location corresponds with temperature measurements taken at the Atlanta Hartsfield-Jackson International Airport. Three future climate scenarios were taken into comparison, SSP245, SSP370, and SSP585, corresponding to low, medium, and high projections of future greenhouse gas emissions and climate change.

The predictions result in a future climate archive, where meteorological information, including temperature, humidity, and wind speed, are provided in 30-meter spatial resolution and hourly bins, for 2019 to 2100.

*Cumulative exposure calculation*

Cumulative exposure for each traveller is based on the second-by-second trajectory throughout their trip. In this study, origin-destination trip information from the Survey are combined with the network configuration provided with open-source General Transit Feeds Specification (GTFS) data [21]. Transit trip modelling is conducted using TransitSim 4.0, a model developed, maintained and regularly updated by the NCST research team at the Georgia Institute of Technology [22–24].

Detailed trip trajectories, combined with the predicted future climate archive, feed into the HeatPath Analyzer, developed by Fan et al. [19]. The HeatPath Analyzer is a Python-based program tailored to model cumulative heat stress and risk level associated with multimodal urban transportation at the trip level. The program includes three modules: the first module develops thermal comfort levels based on United States National Weather Service (NWS)'s Heat Index and Wind Chill equations and provided meteorological data [25,26]. The program generates a second-by-second activity profile for each traveller-trip from the detailed trip trajectory and assigns each second of the trip to the travel activity (e.g., walk access to the transit stop, waiting for the bus to arrive, riding on the bus, etc.). The intensity of travel activity is taken from the Compendium of Physical Activities [27–30], which provides intensity levels that have been derived from epidemiological studies for a variety of activities. Lastly, the program uses United States Centers for Disease Control and Prevention (CDC)'s standards to dynamically quantify cumulative heat exposure and assess associated trip-level risk levels [31,32].

The HeatPath Analyzer processes 246 times, once for each model year and each modelled scenario. The high computational demand for each single model run and the large sample size involved precluded multiple simulations for each scenario and model year; hence, the multi-modal mean projected for each future year is presented [20]. The analysis is conducted using High-Performance Computing (HPC) at the PACE supercomputing cluster at Georgia Tech.

The outputs of the cumulative exposure calculation include two trip-level metrics: a cumulative heat exposure score, interpreted as a rest deficit for each trip (measured in minutes), and a risk level, a binary variable indicating whether the overall exposure during the trip exceeds the occupational heat safety standards set by the CDC's Work/Rest Schedule. Both metrics are recorded for each trip, under each scenario, in each modeled calendar year.

*Equity assessment*

The equity assessment is conducted across five demographic features: rider age, rider race/ethnicity, rider gender, household annual income, and household vehicle ownership. The analyses explore the different number of trips taken by each demographic group and the percentage of trips under extreme heat risk. Due to the strong influence of airport-based trips, all trips with an origin or destination to the airport are excluded from the general demographic analysis.

**RESULTS**

In the base year of 2019, the analytical results indicate that 8.3% of transit trips in the summer are conducted under the risks of extreme heat (i.e., surpassing the occupational safety standards proposed by NIOSH). The following sections provide a detailed breakdown of heat impacts across demographic groups for 2019 (Section 3.1) and under future climate scenario conditions (Section 3.2). Section 3.3 integrates the future prediction and the equity assessments, exploring the conditions faced by different groups under future scenarios. Section 3.4. provides insights on how future climate influences different transit mode segments similarly or differently.

*Cumulative exposure of different demographics*

This analysis examines the data from two perspectives: 1) Specific Exposure Rate, which estimates the percentage of trips taken by each demographic group that are at risk of extreme heat, and 2) General Exposure Composition, which details the distribution of trips across Safe Travel and Heat Risk conditions for each demographic group. The Specific Exposure Rate provides targeted insight into the immediate risks faced by these groups during their daily travels. In contrast, the General Exposure Composition takes a broader approach, assessing the potential benefits each group could gain from investments aimed at improving overall travel conditions, thereby informing resource allocation and distributive justice strategies.

In terms of age (**Figure 2**a), the specific exposure rate is highest among older adults (65+, 9.7%). Other than older adults, there is a general decreasing trend in percent of trips exposed to extreme heat risk as age increases. Strategies to mitigate heat risk throughout the entire system will primarily benefit the young- through middle-age adults (age 18-44), as they make more than 70% of transit trips under heat risk conditions. While older adults make up a relatively small percentage of transit trips, they are at a higher risk of extreme heat. This indicates a need for implementing strategies specifically for older adults.

The percent of trips under extreme heat risk shows a clear difference between low-income and high-income groups. All groups earning less than $75,000 have at least 5% of their trips exposed to heat risk, with some groups, such as those earning $40,000-$49,999 and $50,000-$59,999, experiencing more than 10% of their trips under heat risk. In contrast, higher-income groups earning more than $75,000 have a much lower percentage of trips under risk, ranging from 1% to 3%. (**Figure 2**b).

African Americans are the most frequent transit riders in Atlanta, making more than 70% of the summer transit trips (excludes airport trips). The African American transit riders are highly vulnerable to heat risk, with 8.7% of their trips being made under heat risk conditions, which is the highest among all race groups (**Figure 2**c). Given the high ridership levels, African American transit riders make 79% of all of the trips made under heat risk conditions. Other ethnicity groups that face high vulnerability are American Indian Alaska (6.7%) and Other Races (7.4%), while White / Caucasian (5.3%) and Asian (4.9%) are less frequently influenced by heat risk.

Comparing across genders, males travel slightly more frequently in transit than females under both safe and heat risk conditions (45.9% and 45.9% for females, and 53.9% and 53.5% for males under safe and heat risk conditions). Both females and males have similar heat risk exposure, with around 8% of their trips under heat risk conditions (**Figure 2**d). The data for Other genders only included 33 data points and was excluded from the figure, but this group may also be facing a higher exposure risk.

Vehicle ownership is the factor that shows the strongest difference across groups. The population that owns no vehicles (the "zero-car" group) takes the majority of transit trips under safe heat conditions (81.6%), and the vast majority of trips under heat risk (97.3%). The zero-car group has a notably higher share of trips under heat risk (9.8%) than seen in other vehicle ownership groups (1.3% for one vehicle, and 2.2% for two or more vehicles). Dependent riders are less likely to switch to other options under unfavourable weather conditions and are exposed to higher temperatures than others (**Figure 2**e).

In summary, the equity assessment reveals a pattern that in the study area, resource-constrained groups (older adults, lower income, racial minority, and zero-car populations) are disproportionately influenced by extreme heat during their transit trips. System-level strategies to enhance thermal comfort for all transit users will likely enhance the travel conditions of African Americans and zero-car populations, while older adults and lower-income groups may require more targeted intervention.

*Overview of thermal comfort under future climate*

Under 2019 weather conditions, 8.3% of summer trips and 0.0% of winter trips are estimated to be conducted under extreme heat and wind chill risk. Under all scenarios, wind chill risk in the Atlanta area remains at 0.0%. However, with projected future climate, the percentage of trips under the risk of extreme heat grows steadily at a modest rate. Until 2100, 37.1%, 56.1%, and 76.4% are projected to be under extreme heat risk in scenarios SSP245, SSP370, and SSP585, respectively (Table 1, Figure 3). Because none of the modelled trips are exposed to wind chill risk, the discussions in this and the following sections focus on extreme summer heat exposure.

**Table 1**. Percent of Trips Exposed to Extreme Heat (in Summer)

| Year/Scenario | SSP245 | SSP370 | SSP585 |
|---|---|---|---|
| **2019** | | 8.33% | |
| **2030** | 11.6% | 8.40% | 18.7% |
| **2040** | 16.4% | 19.4% | 23.5% |
| **2050** | 20.1% | 21.7% | 30.6% |
| **2060** | 20.3% | 25.0% | 38.7% |
| **2070** | 24.4% | 32.3% | 48.6% |
| **2080** | 31.9% | 41.3% | 59.5% |
| **2090** | 30.4% | 46.9% | 67.5 % |
| **2100** | 37.1% | 56.1% | 76.4% |

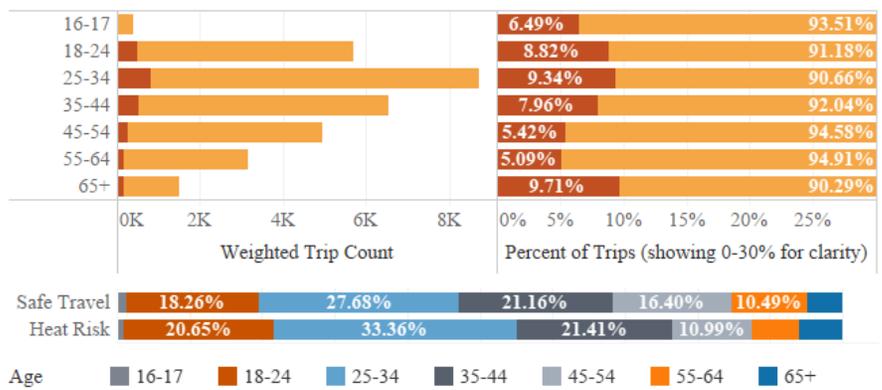
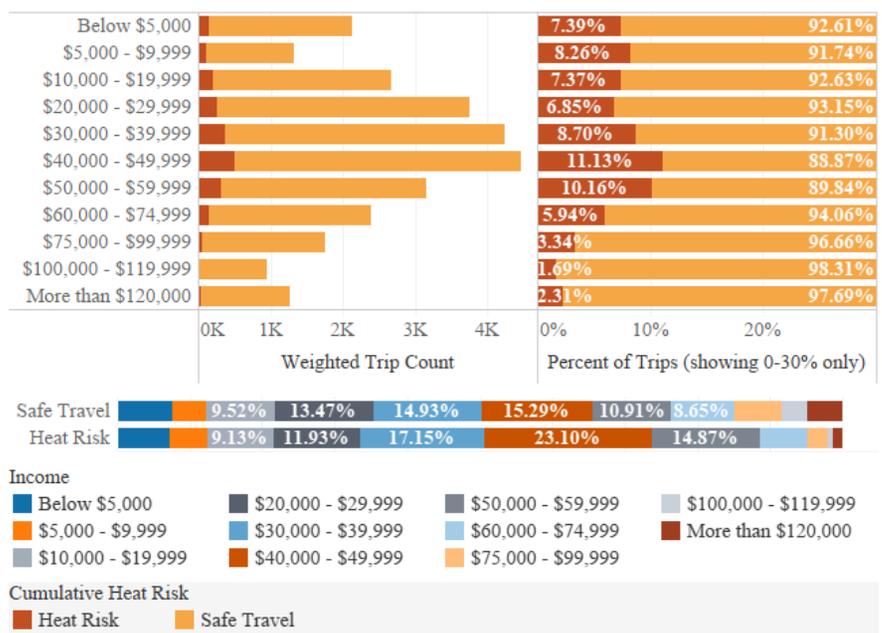
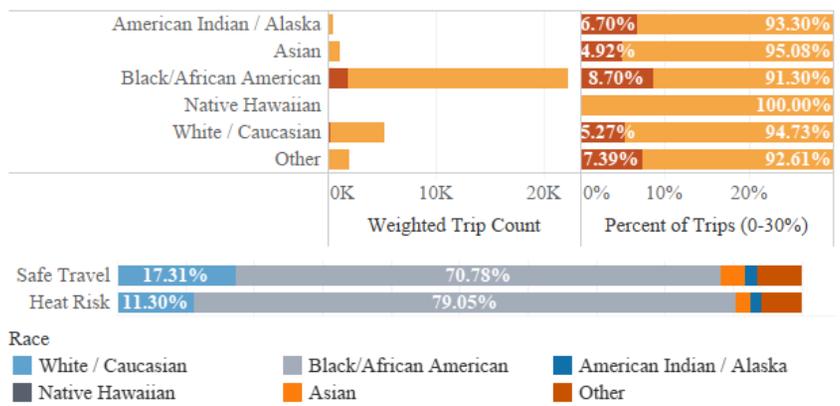
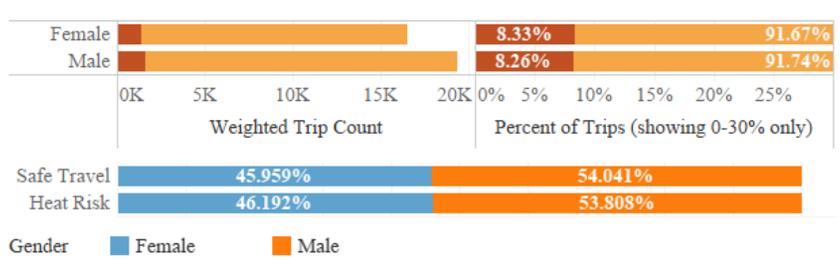
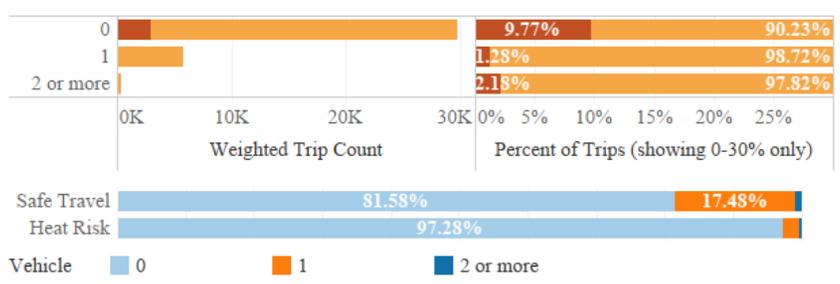

**Figure 2**. Results of Demographic Assessment

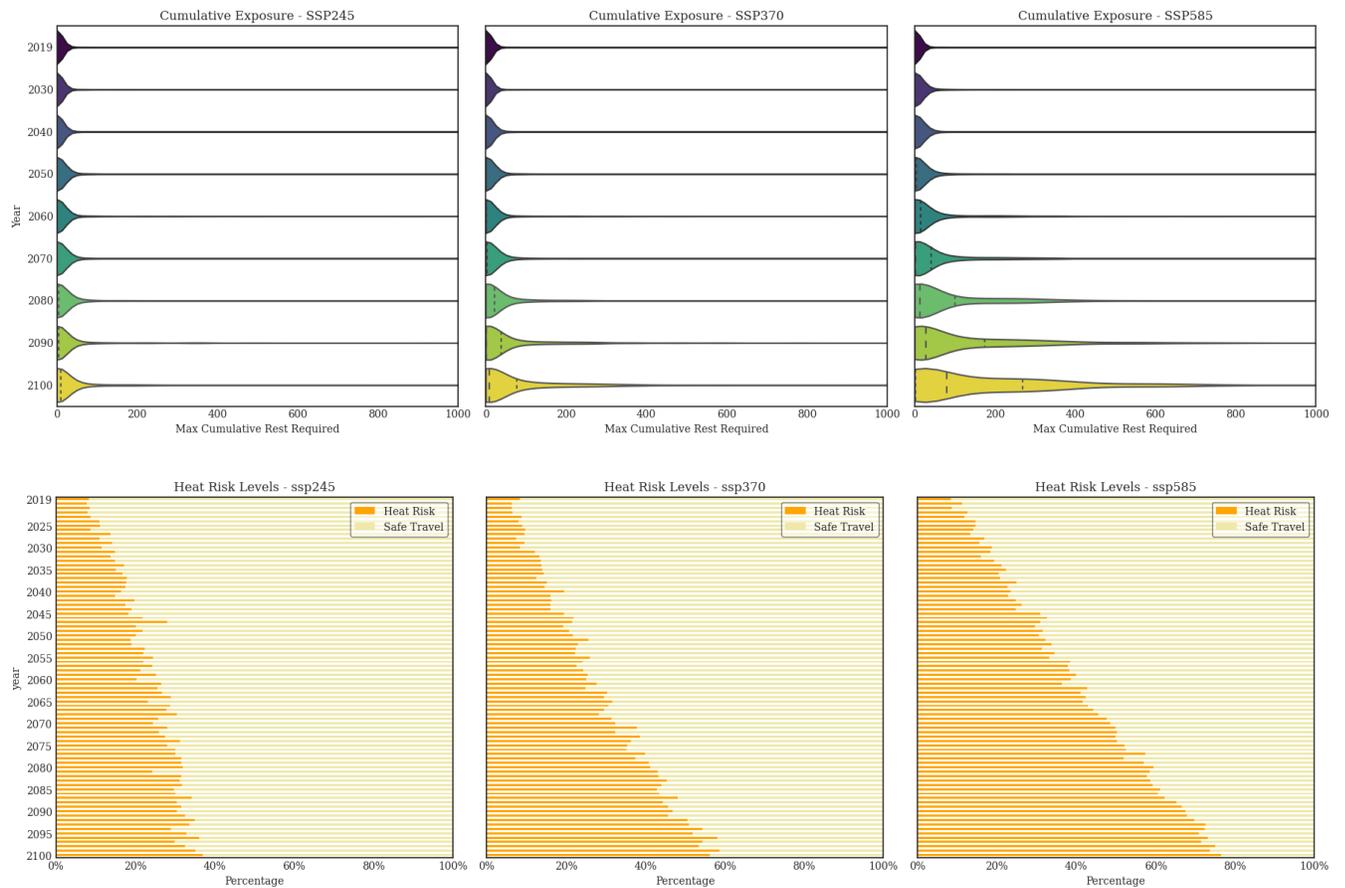

**Figure 3**. Summary of Trip-level Cumulative Exposure (top) and Percent of Trips exposed to Extreme Heat Risk (bottom) by Scenarios and Model

*Future climate impacts across demographic groups*

Repeating the same analyses across the three future climate scenarios for low, medium, and high projections of future greenhouse gas emissions and climate change provides insight into whether the equity impacts identified for 2019 travel may change over time. For different age groups, older adults (65+) and young adults (18-24) are found to have highest percentage of trips under heat risk in summer. All age groups show a steady increase pattern with future projected climate change, with a smaller increase to 33-43% under SSP245 and a larger increase to 73-82% under SSP585 by 2100. (Figure 4a).

The exposure conditions across transit users of different income groups are shown in Figure 4b. The patterns across different groups largely align with the general pattern that is described previously. The lower income groups are disproportionately influenced in the base year, and the disparity is projected to increase over time.

A consistent pattern of a steady increase in heat risk percentage is seen across all race/ethnicity groups. However, Black / African American populations are found to be not only the most heavily exposed at the 2019 baseline, but also experience the steepest increase over time (percent trip exposure grows to 39.0%, 58.7%, and 78.7% under SSP245, SSP370, and SSP585, respectively) (Figure 4c).

The exposure conditions among different genders are shown in Figure 4d. Females and male do not show distinctive exposure over time. The gender "Other" is excluded from this discussion due to the small sample size.

The exposure conditions across travellers with different vehicle ownership are shown in Figure 4e. The zero-car group consistently has the highest percentage of trips exposed to heat risk across all years and under all three scenarios. Additionally, the trend shows that groups with more vehicles experience lower heat risk compared to those with fewer vehicles.

### *Future climate impacts by transit mode segments*

Figure 4f shows the percentage of transit access trip segments (walk, bike, and micromobility), percentage of transit arrival wait time segments, and percentage of walk transfer segments that are exposed to heat risk across future projected times and scenarios. Note that these plots are based on vulnerability-based measurements, meaning that they suggest that traveling on the mode segments involve potential experience of risks, instead of contribution to a total risk. Transit riding and automobile ingress/egress are not included in the discussion because they are typically air-conditioned with low exposure risks.

Walking transfers are the highest exposure element in the base year and have remained either the highest or second-highest across different years and scenarios. This underscores the importance of strategies aimed at reducing the number of walk transfers and walk transfer time. A similar pattern is observed with biking, which consistently maintains a relatively high level compared to other modes. Interestingly, micromobility starts at a relatively low percentage but shows a steep increase under scenarios SSP370 and SSP585, eventually ranking highest by 2100. This is likely due to the long ingress/egress distance associated with micromobility. Considering the high percentage of biking and the projected growth in micromobility, integrating these modes into transit system planning is crucial. They are important elements in enhancing the overall transit rider experience and significantly impacting their health and well-being.

**DISCUSSION**

*Contributions of the current study*

This study seeks to understand thermal comfort for transit riders under rapid urbanization and increasing extreme temperature events. By conducting multi-years simulation using the HeatPath Analyzer and TransitSim 4.0, cumulative heat exposure and trip-level risk is assessed across different demographic groups and their projected future travel conditions. The findings from this analysis show significant disparities in heat risk, particularly affecting vulnerable groups such as zero-car households and ethnic minorities.

The research results also project the impacts of climate change on transit riders under three different future climate scenarios. Providing a comprehensive view of potential future conditions, it helps planners understand how the intensity and frequency of extreme temperature events might evolve and what implications may arise for urban transportation systems. By overlapping future climate scenarios with demographic factors, the study offers insights into how various groups will be disproportionately impacted by climate change in the future. This analysis is crucial for identifying specific vulnerabilities and guiding long-term transit planning.

*Policy implications*

The findings of this study have several important policy implications. Special attention is needed for vulnerable groups such as older adults, zero-vehicle households, and African American transit riders, who are disproportionately affected by extreme heat. For some groups, such as older adults, targeted support is recommended due, to their low share of transit riders and unique health conditions they face. For others, like zero-vehicle households and African Americans, overall improvement of the entire transit network is more effective considering their high percentage of all transit riders.

The results also show that transfer walking is the most challenging segment of current and future transit trips. Therefore, policymakers can prioritize equitable access to heat mitigation resources, such as cooling centres and shaded transit stops, to ensure safety and comfort during transfer. Additionally, biking and micromobility segments may become increasingly vulnerable due to future climate change. Targeted interventions should address active transportation.

Moreover, urban planning and transportation policies need to integrate climate projections to ensure resilience against extreme weather events, creating flexible, adaptive strategies that can respond to varying climate scenarios. As the results show, 37.1 – 76.4% of transit riders will be heavily affected in the future, so investment in green infrastructure, such as urban greenery and permeable surfaces, will be necessary to help reduce the urban heat island effect and improve thermal comfort for all transit users. By doing so, transit agencies can better protect every user, ensure safe and reliable travel, and promote sustainability in urban environments.

*Study Limitations*

This study utilizes land surface temperature (LST) derived from satellite imagery to extrapolate spatial variations in air temperature. This method effectively captures the mesoscale impacts of urban heat islands and greenery cooling effects, which are not as discernible from meteorological station data alone [1]. However, LST data present limitations; LST can sometimes overestimate the magnitude of urban heat island effects and exhibits variability near urban greenery [33,34]. Comparisons of satellite-derived LST with ambient air temperature have shown that despite some discrepancies, LST generally aligns well with air temperature patterns, confirming its utility in representing spatial temperature variations [35,36]. Additionally, cloud cover significantly influences the accuracy of satellite-based LST products, as high cloud cover can skew LST estimates [37]. To minimize this effect, this study employed primarily cloud-free images (cloud cover < 10%), thereby reducing potential biases and more closely reflecting actual air temperature conditions. With projected future climate change, the disparity will be more pronounced.

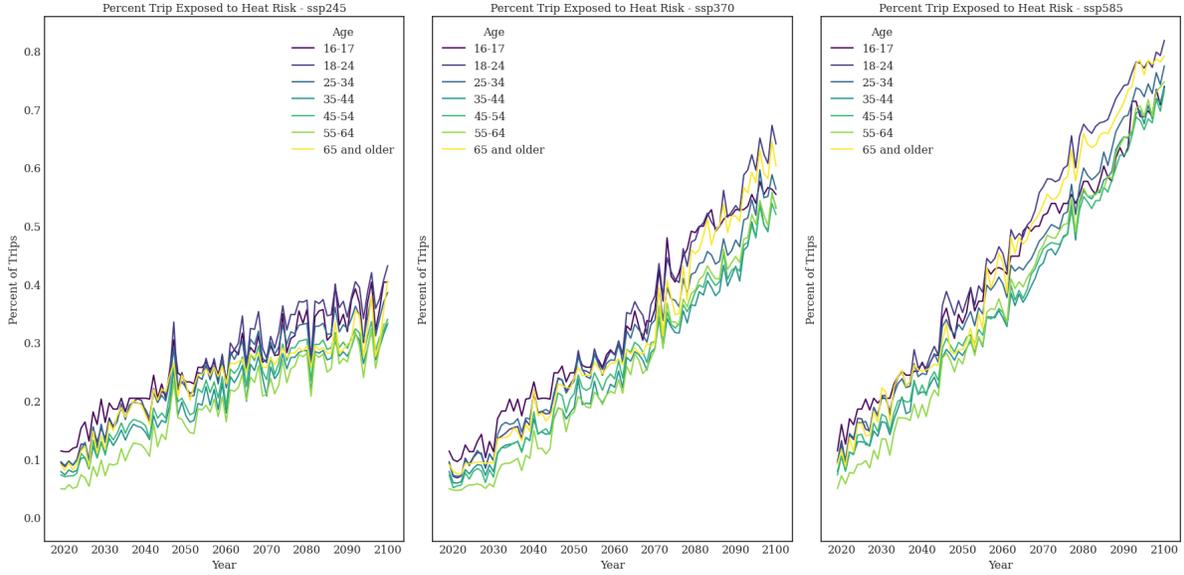

**(a). Age**

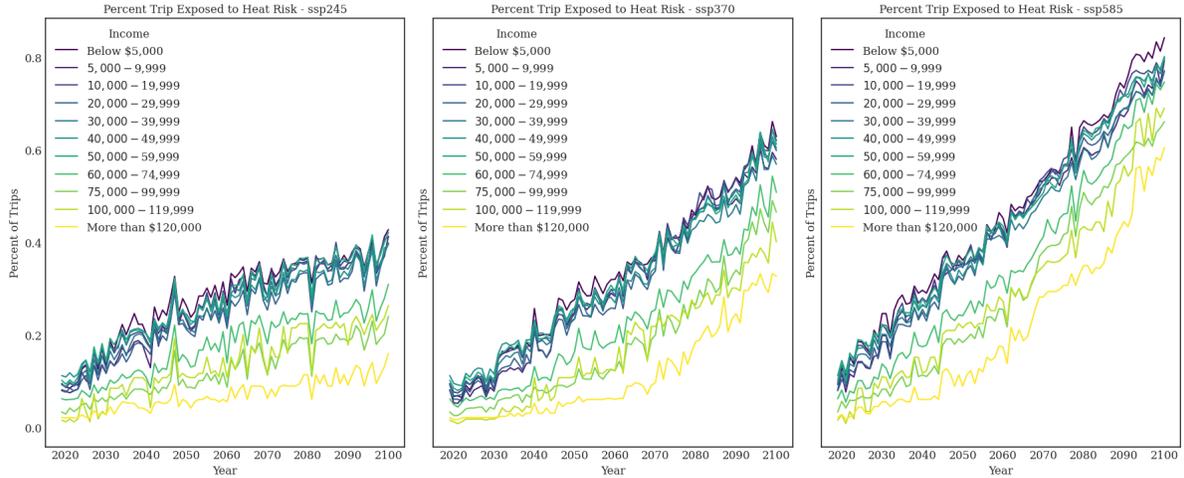

**(b). Income**

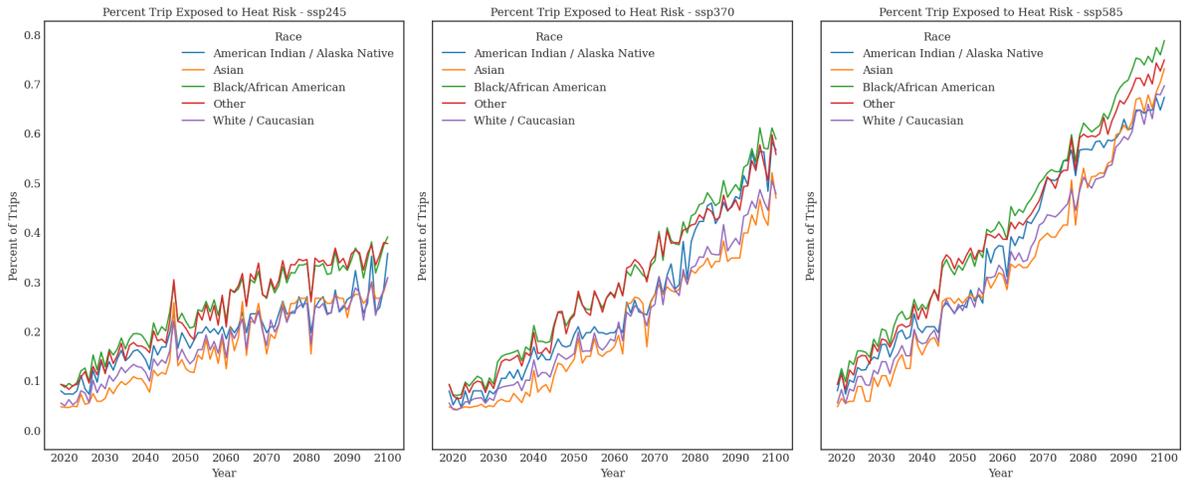

**(c). Race**

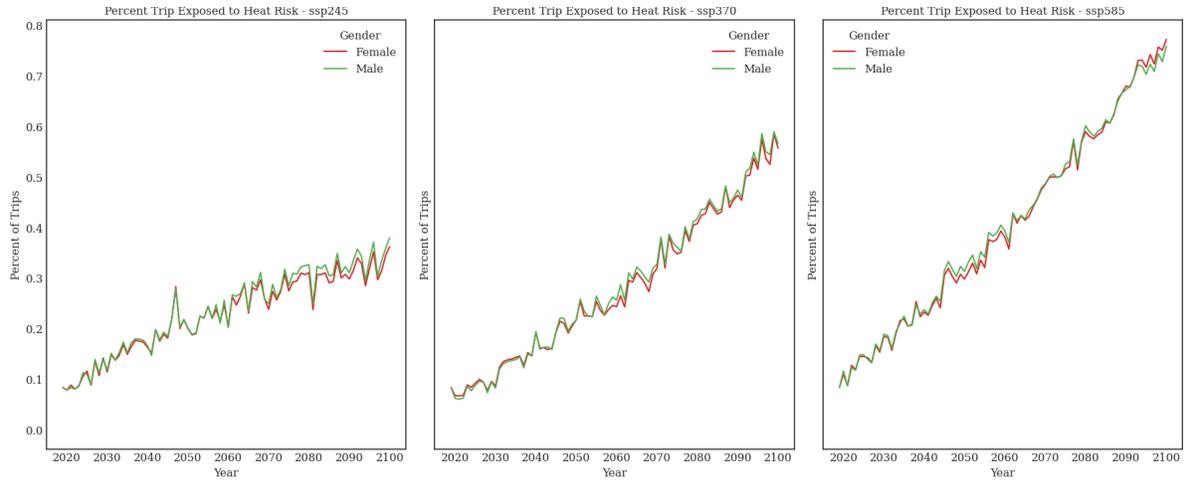
**(d). Gender**

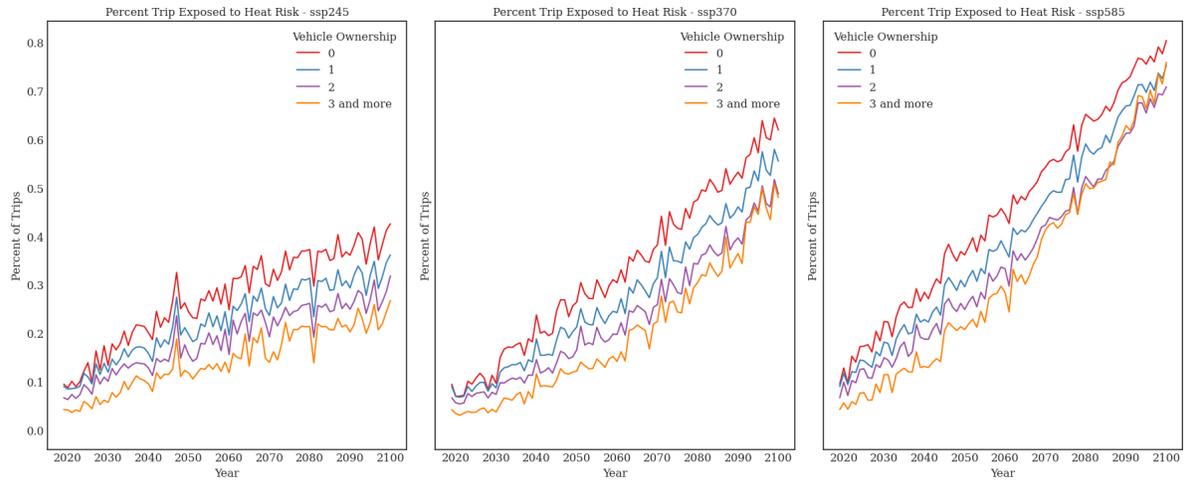
**(e). Vehicle Ownership**

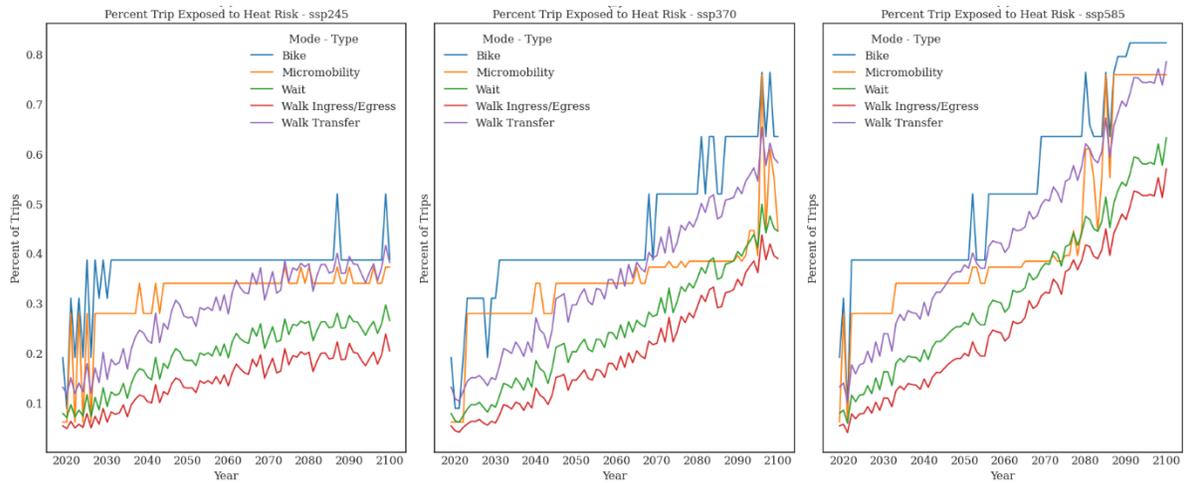
**(f). Transit Trip Component**

**Figure 4.** Percent of Trips Exposed to Heat Risk Over Time by Climate Scenario and Demographic Group (a-e) and Transit Trip Component (f)

# CONCLUSIONS

This study modeled heat exposure across demographic groups under various climate scenarios to predict the potential equity implications of future climate impacts on transit riders. The analysis indicates that 8.33% of summer trips are currently conducted under extreme heat conditions, with this number increasing significantly over time as climate change worsens. By 2100, the proportion of trips at risk of extreme heat is projected to reach 37.1%, 56.1%, and 76.4% under scenarios SSP245, SSP370, and SSP585, respectively. Equity analysis reveals that transit riders that do not own vehicles, that are from low-income households, that are older adults, and that are African Americans are disproportionately affected by extreme heat. This disparity is expected to persist or even widen as climate change progresses. Walking transfers contibute significantly to the trips that at risk for heat exposure. Bike and micromobility access elements are expected to be increasingly affected by heat in the future.

# ACKNOWLEDGMENTS

This study is funded and supported by the National Center for Sustainable Transportation (NCST). The authors also want to thank the Atlanta Regional Commission (ARC) for their long-term activities associated with the continual improvement of the activity-based travel demand model and for their never-ending support of research at Georgia Tech by providing access to model-run outputs and their technical expertise.